\documentstyle[12pt]{article}
\newlength{\extraspace}
\newlength{\extraspaces}
\newcommand{\be}{\begin{equation}
\addtolength{\abovedisplayskip}{\extraspaces}
\addtolength{\belowdisplayskip}{\extraspaces}
\addtolength{\abovedisplayshortskip}{\extraspace}
\addtolength{\belowdisplayshortskip}{\extraspace}}
\newcommand{\ee}{\end{equation}}

\newcommand{\ba}{\begin{eqnarray}
\addtolength{\abovedisplayskip}{\extraspaces}
\addtolength{\belowdisplayskip}{\extraspaces}
\addtolength{\abovedisplayshortskip}{\extraspace}
\addtolength{\belowdisplayshortskip}{\extraspace}}
\newcommand{\ea}{\end{eqnarray}}

\newcommand{\nonu}{\nonumber \\[.5mm]}
\newcommand{\A}{&\!\!\!}

\newcommand{\newsection}[1]{
\vspace{7mm} \pagebreak[3] \addtocounter{section}{1}
\setcounter{subsection}{0} \setcounter{footnote}{0}
\begin{center}
{\large {\bf \thesection. #1}}
\end{center}
\nopagebreak
\medskip
\nopagebreak \hspace{3mm}}

\begin{document}
\pagenumbering{roman}

\newpage

\pagenumbering{arabic}

\begin{center}
{{\bf M\o ller's energy of the Kerr-NUT Metric}}
\end{center}
\centerline{ Gamal G.L. Nashed}

\bigskip

\centerline{{\it Mathematics Department, Faculty of Science, Ain
Shams University, Cairo, Egypt }}

\bigskip
 \centerline{ e-mail:nashed@asunet.shams.edu.eg}

\hspace{2cm}
\\
\\
\\
\\
\\
\\
\\
\\
The energy distribution of the Kerr-NUT space-time is calculated
using M\o ller's energy-momentum complex  within the framework of
the Riemannian geometry.\\

\hspace*{-.4cm}PACS: 04.20.Cv, 04.20.Fy\\
\hspace*{-.1cm}Keywords: Kerr-NUT space-time, M\o ller's
energy-momentum complex.
\begin{center}
\newsection{\bf Introduction}
\end{center}

The search for a consistent expression for the gravitating energy
and angular momentum of a self-gravitating distribution of matter
is undoubtedly a long-standing problem in general relativity (GR).
It is believed that the energy of the gravitational field is not
localizable, i.e., defined in a finite region of the space. The
gravitational field does not possess the proper definition of an
energy momentum tensor and one usually defines some
energy-momentum as Tolamn \cite{To}, Landau-Lifschitz \cite{LL},
Goldberg \cite{GO}, Bergmann \cite{BT},
 M\o ller \cite{Mo5}, Komar \cite{Ko}, Arnowit et al. \cite{Ar} Nahmed-Achar
 and Schutz \cite{NS1}, Bratnik \cite{Ba}, Kataz and Ori \cite{KO} etc. which are
pseudo-tensors and depend on the second derivative of the metric
tensor. These quantities can be annulled by an adequate
transformation of coordinate. They \cite{LL,BT} justify the
results as being consistent with Einstein's principle of
equivalence. In this principle, it can be always find a small
region of space-time, where it prevails Minkowski space-time.  In
such a space-time, energy of the gravitational field is null.
Therefore, it is only possible to define the energy of the
gravitational filed in whole space-time region and not only in a
small region.

M\o ller \cite{Mo5,Mo6} proposed an expression which is the best
to make calculations in any coordinate system within the framework
of  Riemannian geometry. He claimed that his expression would give
the same results for the total energy and momentum as the
Einstein's energy-momentum complex for a closed system. Lessner
\cite{Lg} gave his opinion that M\o ller's definition is a
powerful concept of energy and momentum in (GR). However, M\o ller
prescription was also criticized by some researchers
\cite{Bh}$\sim$\cite{Ka}. Komar's complex \cite{Ka}, though not
restricted to the use of Cartesian coordinate, is not applicable
to non-static space-times. Thus each of these energy-momentum
complex has its own drawbacks. As a result, these ideas of the
energy-momentum complexes could not lead to some unique definition
of energy in (GR).

Several attempts have been made to resolve the problem of
energy-momentum localization but still remains out of hand. This
problem first appeared in electromagnetic which turns out to be a
serious matter in GR due to the non-tensorial quantities.
Virbhadra et al. \cite{RV} explored several space-times for which
different energy-momentum complexes show a high degree of
consistency in giving the same and acceptable energy-momentum
distribution. Aguirregabira et al. \cite{ACV} showed that five
different energy-momentum complexes gave the same result for any
Kerr-Schild class (including the Schwarzschild,
Reissner-Nordstr$\ddot{o}$m, Kerr and Vaidya metrics). An
extension of these calculations has been done
\cite{Xs}$\sim$\cite{R1}.

It is the aim of the present study to calculate the energy,
momentum and angular momentum of the Kerr-NUT space-time using the
energy-momentum complex given by M\o ller \cite{Mo5} within the
framework of GR. In \S 2, we give a brief review  of Kerr-NUT
space-time. Calculation of energy using the energy-momentum
complex  of M\o ller is given in \S 3. A special cases of the
energy is also discussed in \S 3. Final section is devoted to main
results.

\newsection{Kerr-NUT space-time}
The Kerr-NUT (Newman-Unti-Tamburino) space-time \cite{NUT}
describes a stationary axi- symmetric object with gravitomagnetic
monopole and dipole moments associated with nonzero values of the
NUT and Kerr parameters $L$ and $a$  respectively and as such a
useful model for exploring gravitomagnetism \cite{BC}. The
Kerr-NUT space-time and its spacial  cases all belong to the
larger class of stationary axi-symmetric type D vacuum solutions
of the Einstein equations found by Carter \cite{Cb} for which the
Hamilton-Jacobi equation for the geodesic is separable. The
stability of the Kerr-NUT space-time is probed  by studying their
perturbation by fields of various spin \cite{BD,BDC}. The Kerr-NUT
metric in Boyer-Lindquist-like coordinates is given by the
following line-element \ba ds^2=\A \A \frac{1}{\Sigma}
 \left\{\Delta-a^2\sin^2 \theta  \right\} dt^2 -\frac{\Sigma}{\Delta} d\rho^2
  -\Sigma d\theta^2- \frac{1}{\Sigma}\left\{ (\Sigma+a
\chi)^2 \sin^2\theta-\chi^2 \Delta
 \right\} d\phi^2\nonu
\A \A -\frac{2}{\Sigma} \left\{\chi \Delta-a(\Sigma+a\chi) \sin^2
\theta \right\} dt d\phi
 \; ,\ea
 and the corresponding electromagnetic Faraday tensor can be
 expressed in terms of the 2-form
 \ba {\bf
 F} \A=\A \frac{Q}{\Sigma^2}\Biggl[\left(\rho^2-\{l+a\cos\theta\}^2\right)d\rho
 \wedge (dt-\chi d\phi)\nonu
 \A \A+ 2\rho \sin\theta(l+a\cos\theta)d\theta \wedge
 \left(\{\rho^2+a^2+l^2\}d\phi-a dt\right)\Biggr],\ea
 where $\Sigma, \ \  \Delta, \ \   and \ \ \chi$ are defined by
  \be \Sigma = \rho^2+(l+a \cos\theta)^2\; ,  \qquad  \Delta=\rho^2-2M\rho-l^2+a^2+Q^2
  \; , \qquad \chi=a\sin^2\theta-2l\cos\theta\;. \ee
  Units are chosen such that $G=c=1$, so that $(M, Q, a, l)$
  all have the dimension of length: the source has  mass M, electric charge Q,
   angular momentum $J=Ma$ (i.e., gravitomagnetic dipole moment)
   along the z-direction, and gravitomagnetic monopole moment
   $\mu=-l$, where $l$ is the NUT parameter.
\newsection{M\o ller energy-momentum complex}
M\o ller's energy-momentum complex is given by \be
{\jmath_\mu}^\nu=\frac{1}{\kappa} {\chi_\mu}^{\nu \rho}_{,\rho},
\ee satisfying the local conservation laws \be \frac{\partial
{\jmath_\mu}^\nu}{\partial x^\nu}=0, \ee where the antisymmetric
superpotential  $ {\chi_\mu}^{\nu \rho}$ is defined by \be
 {\chi_\mu}^{\nu \rho}=-{\chi_\mu}^{\rho \nu}\stackrel{\rm def.} {=}
 \sqrt{-g}\left(g_{\mu \delta,\gamma}-g_{\mu
 \gamma,\delta}\right)g^{\nu \delta}g^{\rho \gamma}.\ee The energy
 and momentum components are given by \be P_\mu=\int \int \int
 {\jmath_\mu}^0 dx^1dx^2dx^3, \ee where $P_0$ is the energy while
 $P_\alpha$ stand for the momentum components. Using the Gauss's
 theorem the energy $E$ for a stationary metric is given by \be
 E=\frac{1}{8\pi} \int \int {\chi_0}^{0 \alpha} n_\alpha dS,\ee
 where $n_\alpha$ is the outward unit normal vector over an
 infinitesimal surface element ${\it dS}$.

 As is clear from Eq. (8) that the only required component is ${\chi_0}^{0 1}
 $. Using Eq. (6) in Eq. (1) we get
 \be {\chi_0}^{0 1}={2 \sin\theta (\rho^2+a^2+l^2)
\over \{\rho^2+(a\cos\theta+l)^2 \}^2} \Biggl[M\rho^2-\rho
Q^2-Ma^2\cos^2\theta-Ml^2+2\rho l^2+2\rho a l \cos\theta \Biggr],
\ee using Eq. (9) in Eq. (8) we get \be E_{M\o l}  =
M+\displaystyle{2l^2 \over \rho} -\displaystyle{Q^2 \over 2
\rho}\left[1+\left\{\displaystyle{ (a^2+l^2+\rho^2) \over 2 a
\rho}-{a \ l^2 \over \rho Q^2} \right\} \left(\tan^{-1}
\displaystyle{\left\{a-l \over \rho \right\}}+\tan^{-1}
\displaystyle{\left\{a+l \over \rho \right\}}\right)\right] \nonu
.\ee The formula of energy given by Eq. (10) is a general formula
and reduces to:\\
i) The Kerr-Newman energy when the NUT parameter $l=0$ \cite{Xs,
NS}. \vspace{0.5cm}\\
ii) The Kerr energy when $Q=l=0$ \cite{NS}.\vspace{0.5cm}\\
iii) The Reissner-Nordstr$\ddot{o}$m energy when $a=l=0$
\cite{Nma}.\vspace{0.5cm}\\
iv) The Schwarzschild energy when $a=l=Q=0$ \cite{Nsp}. \vspace{0.5cm}\\

\newsection{Main results and discussion}

In GR, several attempts to obtain a meaningful and generally
covariant expression for the energy density have proven
unsuccessful \cite{AH}. The canonical energy-momentum pseudotensor
expression derived from variational formulation of general
relativity has been found neither unique nor tensorial. Although
Komar \cite{Ko} and Penrose \cite{Pr} have developed some
sophisticated approach, the canonical energy-momentum
pseudotensors have been found useful in the computation of
integrated energy, provided some restriction is imposed. Komar
\cite{Ko} definition is not restricted to "Cartesian coordinate"
like one has for the Tolamn \cite{To}, Landau-Lifschitz \cite{LL},
Goldberg \cite{GO}, Bergmann \cite{BT}. However, Komar definition
is applicable only to stationary space-times. The M\o ller
energy-momentum complex is neither restricted to the use of
particular coordinates nor to the stationary space-times. Lessner
\cite{Lg} pointed out that M\o ller definition is a powerful
representation of energy and momentum in GR.

For our background metric given by Eq. (1) we have examined M\o
ller pseudotensor. Then the total energy associated with metric
given by Eq. (1) contained in a sphere of radius $\rho$ can be
found as given by Eq. (10). This calculation shows that the total
energy associated with the space-time given by Eq. (1) is shared
by its exterior as well as interior. In the case of Kerr black
hole, i.e., case (ii) above, it is clear from Eq. (10) that there
is no energy contained by the exterior of the Kerr black hole and
hence the entire energy is confined with its interior. This result
is quite in conformity with that obtained before \cite{RV, NS,AH}.

\end{document}